\documentstyle[prl,twocolumn,aps,epsfig]{revtex}  
\begin{document}                
\title{Photoionization of ultracold and Bose-Einstein condensed Rb atoms}
\author{D. Ciampini \and M. Anderlini \and J.H. M\"{u}ller \and F.
Fuso \and O. Morsch \and
J.W. Thomsen\thanks{Permanent address:
Niels Bohr Institute, Oersted Laboratory, University of Copenhagen,
DK-2100 Copenhagen,
Denmark.} \and and E. Arimondo.}
\address{INFM, Dipartimento di Fisica E. Fermi, Universit\`{a} di Pisa, Via
Buonarroti 2, I-56127 Pisa, Italy}

%
\maketitle

\begin{abstract}                

Photoionization of a cold atomic sample offers intriguing
possibilities to observe collective effects at extremely low
temperatures. Irradiation of a rubidium condensate and of cold rubidium
atoms within a magneto-optical trap with laser pulses ionizing
through 1-photon and 2-photon absorption processes has been
performed. Losses and modifications in the density profile of
the remaining trapped cold cloud or the remaining condensate
sample have been examined as function of the ionizing laser
parameters. Ionization cross-sections were measured for atoms in a
MOT, while in magnetic traps losses larger than those expected for
ionization process were measured.
\end{abstract}
\section{Introduction}               
The recent development of laser cooling and trapping techniques
has made possible the controlled realization of dense and cold
atomic samples, thus opening the way for spectroscopic
investigations in the low and ultra-low temperature regimes not
accessible with conventional techniques. Cold atoms in
magneto-optical traps (MOT) have been used as excellent tools for
the measurement of ionization cross sections. In particular for
alkaline
atoms\cite{dinneen,gabbanini,gabbanini2,cesium,knize,fuso,graz,duncan}
and for magnesium\cite{magnesium} valuable data have been obtained
using the trap population dynamics as a sensitive monitor of the
trap loss induced by the ionizing radiation, extracting wavelength
dependencies and absolute values for the cross sections. More
recently ultra-cold plasmas have been created making use of laser
cooled atomic samples and laser excitation to high Rydberg
states\cite{plasma}.  To understand the relaxation of such a
system towards equilibrium many phenomena such as recombination
and super elastic collisions in a completely new temperature
regime for plasmas have to be taken into account. Moreover  a
frozen Rydberg gas can be created with intriguing transport
properties, like quasi-metallic behavior  or delayed electron
emission \cite{cote}. In these systems the collective character of
the dynamics is due to the strong interaction between charged and
highly excited particles.

The realization of Bose-Einstein condensates (BEC) of alkali atom
vapors has attracted much interest to new aspects of photon-matter
interaction arising from the coherent nature of the atomic
ensemble. Recently attention has been paid to the analysis of
photoionization of a BEC by monochromatic laser light
\cite{mazets}. The products of the photoionization process
(electrons and ions) obey Fermi-Dirac statistics. Owing to the
coherent nature of the initial atomic ensemble and the narrow
spectral width of the laser ionization source, the  occupation
number of the electron/ion final states can approach unity,
especially for excitation close to threshold. In this regime the
ionization rate should be reduced by the Pauli blockade and
determined by a balance between the laser light acting on the
atoms and the rate of escape of the ionization products from the
condensed system.

When photoionizing cold atoms, at decreasing temperature
and at increasing atomic density, the interaction between the
charged particles and the neutral "dielectric"
background can become significant. New information about the
energy transfer from the charged species to the neutral particles
can be obtained using non-dissipative traps, such as magnetic
traps, and using the atom dynamics as a monitor. An ion localized
inside a condensate modifies the density distribution of the
condensate dielectric background by attracting the atoms towards
itself. The classical interaction energy between a single ion and
a polarized ground state condensate at a distance typical for a
condensate density of $10^{14}$ cm$^{-3}$ is larger than typical
chemical potentials in Bose-Einstein condensates ($h$ times 100
Hz). This means that locally the polarization effect is important
for the condensate dynamics and may even lead to a destruction of
the condensate phase. As a consequence, the observable effect
depends on the interaction time between ions and condensate: apart
from the recoil picked up from the ionizing photon, in real
experiments stray electric fields might limit this interaction
time. Examining the condensate cloud remaining after the laser
excitation,  the atom-ion/electron interaction for weakly ionized
clouds may show significant changes of the density and momentum
distribution\cite{helium}. Furthermore the interaction of
the charged particles among themselves will  play an important
role for higher fractions of ionized particles, with plasma
formation and collective effects.

In this paper we report, and compare, the results for the
experimental investigation of the photoionization process within a
rubidium condensate and within the cold rubidium sample of a MOT.
The condensate photoionization process originates from the ground
state, while for the MOT sample results from both the ground and
the first excited electronic state have been obtained. We
investigated ionization by means of 1-photon and 2-photon single
color processes using various pulsed and cw laser sources, more
precisely 296 nm for 1-photon ionization and 580-600 nm for 2-photon ionization
from the  5S$_{1/2}$ ground  state,  296 and 421 nm  radiations for
1-photon ionization from the 5P$_{3/2}$ state. The
ionization process has been examined through the losses from the
sample, a cold cloud inside a MOT or a magnetic trap, or a
condensate inside a magnetic trap. Our approach is based on trap
loss ionization spectroscopy with the photoionization losses
measured as an additional decay after the trap loading.

Section II recalls briefly the model for photoionization induced
trap loss, concentrating on the use of a pulsed laser. Section
III describes the experimental set-up and the characteristics of
the applied pulsed and cw laser radiation. Section IV analyzes and
discusses the experimental results obtained on the thermal and
condensate cloud, separating the MOT and magnetic trap
experiments. Section V concludes the present investigation.

\section{Trap loss spectroscopy}
\subsection{Photoionization}
The dynamics of the number $N$ of trapped atoms in a MOT in the
constant density regime and with the trap loading shut off,
predicts an exponential decay, with a time constant $1/\gamma$,
determined by collisions with the background gas and intra-trap
collisions. In the presence of ionization by a cw laser, additional
losses with rate $\gamma_{\rm ph}$ shorten the lifetime of the
trap and the equation of evolution for $N$ becomes\cite{dinneen}
\begin{equation}
\frac{dN}{dt}= -\gamma N -\gamma_{\rm ph} N.
\label{eq:MOTdecay}
\end{equation}
Hence, if $N_0$ is the initial number of atoms,
\begin{equation}
N(t)= N_0 e^{-(\gamma +\gamma_{\rm ph}) t}.
\label{eq:MOTdecaybis}
\end{equation}
When using photons above the ground state ionization threshold  on a
MOT, the loss rate $\gamma_{\rm ph}$  contains the
contributions from the 5S rubidium ground state ionization
rate ${\sf R}_{5S}$ and from the 5P rubidium excited state ionization rate
${\sf R}_{5P}$
\begin{equation}
\gamma_{ph} = (1-f) {\sf R}_{5S} +f {\sf R}_{5P},
\label{gamma}
\end{equation}
with $f$ the fraction of atoms in the excited state. That fraction
$f$, under standard MOT operating conditions, can be calculated
from the MOT laser parameters as in refs.\cite{gabbanini,cesium}.
For a cold cloud in a magnetic trap or for a condensate, Eq.
(\ref{eq:MOTdecay}) can be also applied. In that case the excited
state fraction $f$ is equal to zero, so that the ionization loss
depends on the ground state rate ${\sf R}_{5S}$ only.

In the case of 1-photon ionization by a laser with a photon flux
$F_{\rm ph}$, the ionization rate is
\begin{equation}
{\sf R}_{5l} = \sigma^{(1)}_{5l} \beta^{(1)} F_{\rm ph},
\label{rate}
\end{equation}
with $l=(S,P)$, $\sigma^{(1)}_{5l}$ being the 1-photon ionization
cross-section and $\beta^{(1)}$ a geometrical correction
coefficient. For a 2-photon ionization process with cross-section
$\sigma^{(2)}_{5l}$ and geometrical correction coefficient
$\beta^{(2)}$, the ionization rate is
\begin{equation}
{\sf R}_{5l}= \sigma^{(2)}_{5l} \beta^{(2)} (F_{\rm ph})^{2},
\end{equation}
The geometrical coefficients $\beta^{(k)}$  take into account the
spatial distribution of the  laser beam and the atomic sample.
For a Gaussian laser beam with waists $w_x$ and $w_y$ centered on
a Gaussian atomic distribution with sizes $L_x$ and $L_y$, and
assuming the density distribution of the target atoms not depleted
by the laser, the geometrical corrections are
\begin{equation}
\beta^{(k)} = \frac{1}{\sqrt{\left (1+ k \left
(\frac{L_x}{w_x}\right)^2 \right)\left (1+ k \left
(\frac{L_y}{w_y}\right)^2 \right)} },
\end{equation}
with $k=(1,2)$. The correction approaches unity for laser beam
sizes larger than the atomic sample size. The ionization photon
flux $F_{\rm ph}$ is connected to the photon number $n_{\rm ph}$
and area $A_{\rm ph}=\pi w_x w_y/2$, or to the intensity $I_{\rm
ph}$ and wavelength $\lambda_{\rm ph}$ by
\begin{equation}
F_{\rm ph}= \frac{n_{\rm ph}}{A_{\rm ph}}=\frac{I_{\rm
ph}\lambda_{\rm ph}}{hc}.
\end{equation}

For the ionization by a pulsed laser with pulse duration
$\tau_{\rm ph}$, the ionization probability is
\begin{equation}
P_{\rm ph} = \gamma_{\rm ph} \tau_{\rm ph}.
\label{eq:single}
\end{equation}
In analogy with Eq. (\ref{eq:MOTdecaybis}), the atom number left
in the trap (either magneto-optical or magnetic) after the
application of a single laser pulse is
\begin{equation}
N_{\rm ph}= N_0 (1-P_{\rm ph}) = N_0 (1- \gamma_{\rm ph} \tau_{\rm ph}).
\end{equation}
For a sequence of laser pulses, we suppose that each ionization
process is not modified by the previous photoionization history.
Thus the application of a sequence of $m$ pulses leads to the
following remaining number:
\begin{equation}
N_{\rm ph}= N_0 (1- \gamma_{\rm ph} \tau_{\rm ph})^m.
\label{singlepulse}
\end{equation}
If we  apply a pulse sequence with rate $r_{\rm ph}$, at the
time $t$ the atom number becomes
\begin{equation}
N(t)= N_0 e^{- \gamma t} (1- \gamma_{\rm ph} \tau_{\rm ph})^{r_{\rm ph}t}
\label{pulses}
\end{equation}
In the case of a small ionization probability, the decrease in the
atom number is well approximated by an exponential decay, as in
Eq. (\ref{eq:MOTdecaybis}) for cw laser, with an effective
ionization decay $\gamma^{\rm eff}_{\rm ph}$
\begin{equation}
\gamma^{\rm eff}_{\rm ph}=\gamma_{\rm ph}\tau_{\rm ph} r_{\rm ph} =
\left[(1-f) {\sf R}_{5S} +f {\sf R}_{5P}\right]\tau_{\rm ph} r_{\rm ph}.
\label{effrate}
\end{equation}
We verified Eq. (\ref{pulses}) to be valid by changing the
repetition rate $r_{\rm ph}$, and hence the time separation
between successive pulses, and obtaining consistent results for
$\gamma^{\rm eff}_{\rm ph}$.

\subsection{Scattering and dipole force}
Apart from changing the internal atomic state, the ionizing laser
radiation also acts on the center-of-mass motion of the atoms
through the scattering and dipole forces (see e.g.\cite{metcalf}
for an overview and \cite{meschede} for the case of pulsed
radiation). For ionization above threshold the excess energy of
the photon is converted into kinetic energy of the fragments
imparting large momentum on the electron and ion. For
trap loss spectroscopy only the radiation forces xing on the
atoms remaining in bound states within the trap need to be
considered for additional loss mechanisms. The dissipative
environment of a MOT with a typical effective trap depth of 300 mK
safely recaptures and cools atoms with velocities of some 10 m/s,
while in a conservative magnetic trap with typical depth of 50
$\mu$K atoms accelerated to 10 cm/s can escape from the
trap.

The relative role of the scattering and dipole forces acting on
the atoms depends mainly on the laser detuning from the atomic
transitions. For most of our investigations with pulsed lasers,
the laser detuning was so large that the scattering forces could
be neglected. For the case of a condensate there is a subtle
difference between the effects of the scattering and dipole
forces.  In a  spontaneous emission event  the photon recoil is
picked up by this
atom, as long as the condensate sound velocity is smaller than the
atomic recoil velocity. Instead, the dipole force due to the
coherent redistribution of laser photons acts on the whole
condensate cloud. For a constant intensity gradient over the
condensate, the center-of-mass motion of the condensate will be
driven by the dipole force without internal excitation. By
contrast, for laser beam sizes comparable to the condensate size,
the gradient of dipole force produces also internal excitations of
the condensate. For the impulse approximation in the case of
pulsed radiation, a complicated phase pattern is imprinted onto
the condensate cloud, which subsequently evolves into considerable
modifications of the condensate density distribution, catalyzing
the formation of solitons, as in ref.\cite{burger}, or splitting
the condensate cloud in pieces. Part of these fragments may
acquire a high enough kinetic energy to leave the magnetic trap
potential, hence leading to additional trap losses. Rather than
attempting to model the complicated cloud dynamics, in Section IVb
we will only estimate the kinetic energy imparted by the laser
pulse on a single atom.

\subsection{Collisions}
Additional trap losses may be caused by inelastic collisions
between the ionization products, electrons and ions, and the
thermal or condensed cloud remaining in the trap. We have found
  very little information available for those very low energy
collisional processes.  Using the cross-section for collisions of
10 meV electrons with carbon dioxide molecules\cite{field} or the
scattering length for the electron-Rb collisions reported in
ref.\cite{greene}, we estimated the collisional processes to
produce negligible additional  trap losses. We should
also consider that, after the ionization, a recombination process
produces atoms in highly excited Rydberg states, and those atoms
will thermalize within the condensate. During this process,
inelastic collisions can lead to additional losses, whose contribution
we cannot easily estimate.

\section{Experimental setup}
Our experimental apparatus, described in ref.\cite{jphysbpaper},
was based on a double MOT system, divided into a high vacuum
region optimized for collecting Rb atoms in a MOT and a second
region of low background pressure, into which the atoms were
collected for transfer  to the high vacuum MOT. Photoionization
investigations were performed in the high vacuum MOT. For the
experimental parameters of our MOT lasers (intensity $I=16.5$
mW/cm$^2$, detuning $\delta = -2.2 \Gamma$) we derived an excited
state fraction $f=0.14 \pm 0.02$.

The atoms from the high vacuum MOT can be transferred into a
magnetic trap, in our case a triaxial time-orbiting potential trap
(TOP). Compressing the trapped cloud and applying forced
evaporative cooling with rf induced spin flips, the BEC phase
transition was reached with typically $5\cdot10^{4}$ atoms. "Pure"
$\mid F=2,m_F=2 \rangle $ condensates  contained up to
$2\cdot10^{4}$ atoms in an ellipsoid whose average dimension was in
the 5 $\mu$m range. Instead  the dimension of the noncondensed cloud inside the
TOP was in the 40-100 $\mu$m range. Depending on the chosen sequence of cooling
and trapping steps, noncondensed or condensed  samples with
temperatures ranging from a few hundreds of $\mu$K down to tens of
nK with atomic densities of 10$^{10}$-10$^{14}$ atoms/cm$^{3}$
were prepared.

The present experimental setup does not include a charge detector
to monitor directly the production of positive and negative
charges. Thus the action of the photoionizing laser was monitored
through the decrease of the atoms remaining in the trap, applying
a shadow imaging detection using a 780 nm near-resonant probe laser
beam. The absorptive shadow cast by the cold atoms was imaged onto
a CCD camera. Measurements on both condensates and very cold
thermal clouds were performed after a few milliseconds of free
fall of the released atomic clouds, when their typical dimensions
were of the order of 30-300 $\mu$m. The atomic samples remaining
after the photoionization process contained a number of atoms
large enough to measure their temperature and density
distribution. Within the magnetic trap both a sample of thermal
atoms with about $10^6$ rubidium atoms cooled by evaporative
cooling down to T=400 nK and a condensate were photoionized.

The 4.177 eV energy threshold for ionization of rubidium atoms
from the $5S_{1/2}$ ground state can be reached by a single photon
of 296.815 nm wavelength. Alternatively the ionization threshold
can be crossed using two photons aroud 594 nm, with a
substantially larger radiation intensity needed since there is no
resonant intermediate level for the 2-photon process. We
irradiated the cold atom with pulsed lasers  operating around 594
nm and 296 nm, and with a cw laser at 421 nm.

The 580-600 nm radiation was produced by an excimer-pumped dye laser,
with intensities in the hundreds of MW/cm$^{2}$ range and a laser
pulse duration $\tau_{\rm ph}$ of 10 ns. Laser pulse repetition
rates $r_{\rm ph}$ up to 12 Hz were used.  The excess energy $E_{\rm
ex}$ of
the ionizing photons, and therefore
the kinetic energy of the released electrons, was around  8.5 meV
with an energy spread due to the laser linewidth of around
0.1 meV. The 296 nm radiation was
generated from the yellow radiation by frequency doubling within a
BBO crystal, with 15$\%$ efficiency.  The wavelength used
corresponded to an electron excess energy $E_{\rm ex}$ of 10.4$\pm$0.1
meV. The 594 nm dye laser radiation was focused down to 80 $\mu$m on the
rubidium atoms,
while for the 296 nm radiation, owing to beam astigmatism, beam
waists of respectively $0.36\times1.6$ mm for the MOT  and
$0.16\times0.78$ mm for the TOP were achieved.

For the 421 nm radiation a grating stabilized diode laser
operating at 842 nm injected a tapered amplifier delivering power
in excess of 500 mW. The IR radiation was frequency doubled using
an LBO crystal placed inside an external enhancement
resonator\cite{toptica}. Around 10 mW of narrowband cw radiation
at 421 nm were available. The
421 nm wavelength was chosen in order to use the 6P$_{1/2}$ or
6P$_{3/2}$ levels as near resonant intermediate steps for a
2-photon process from the ground state leading to an
electron excess energy of 1.7 eV. When applied to atoms stored in
a MOT, the 1-photon process out of the populated 5P$_{3/2}$ is
dominant, releasing electrons with an excess energy of 0.34 eV.

\section{Photoionization results}
\subsection{MOT}

\begin{figure}
\centering \psfig{figure=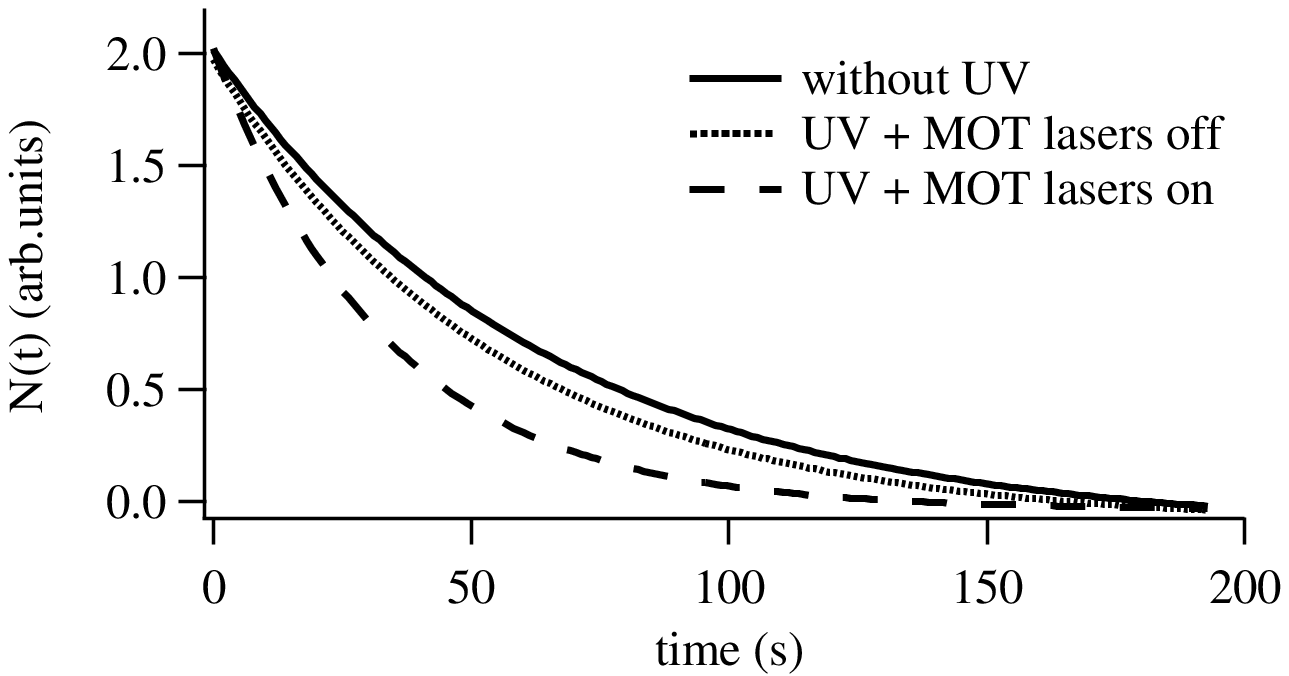, width=8cm} \caption{Time decay
for the fluorescence from the rubidium atom MOT  in absence
(continuous line) and in the presence of  296 nm uv light (dashed
and dotted lines). For the case of the dotted line the MOT lasers
were switched off before applying the uv light in order to
populate the 5S$_{1/2}$ ground state only.  A $r_{\rm ph}=10$ Hz
pulse repetition rate was used, with 130 $\mu$J pulse energy.}
\label{fig:1}
\end{figure}

The frequency doubled radiation 296 nm  pulsed laser induced
single-photon ionization of the ground state and first excited
state.  The MOT atoms were illuminated by a photon flux  $F_{\rm
ph}= 2 \times 10^{24}$ cm$^{-2}$ s$^{-1}$. Decay profiles of the 780
nm wavelength MOT fluorescence,
proportional to the trapped atom number, after shutting of the
loading are displayed  in Fig.\ref{fig:1}. To separate in Eq.
(\ref{effrate}) the  contributions due to the ionization from the
ground 5S$_{1/2}$ and excited 5P$_{3/2}$ state, we combined fast
switching  of the MOT laser beams with the timing of the
ionization pulses. The solid line in Fig.\ref{fig:1} shows the
exponential decay in absence of uv photoionization light, with a
decay time $\gamma^{-1}=$60 s.  The dashed and dotted lines
correspond to the application of the pulsed uv light to a cw and a
synchronously switched MOT, respectively. In the synchronous
operation mode the MOT lasers were switched off for $60 \mu s$
leaving ample time for the atoms to decay back to the ground state
before interacting with the uv pulse while keeping negligible the
ballistic expansion of cloud. Using the fraction of excited atoms
$f$, we derived from $\gamma_{\rm ph}$ (around one fifth of
$\gamma$) the following values of the photoionization
cross-sections:
\begin{mathletters}
\begin{eqnarray}
\sigma^{(1)}_{5 {\rm S}_{1/2}} & = & (0.76 \pm 0.15) \times 10^{-19}
{\rm cm}^2,\\
\sigma^{(1)}_{5{\rm P}_{3/2}} & = & (5.4\pm 1.2) \times 10^{-18} {\rm cm}^2.
\end{eqnarray}
\end{mathletters}
The ground state ionization cross-section value is in reasonable
agreement with  the value  measured
in the late sixties by Marr and Creek\cite{marr} using the
absorption of uv light emitted from a discharge lamp by hot Rb
vapors, and later derived theoretically in refs.
\cite{aymar,fink}.  Since we used interleaved scans for these
measurements, our largest systematic error source, the intensity
determination of the ionizing laser, drops out for the ratio of
the two cross sections
\begin{equation}
\frac{\sigma_{5{\rm P}_{3/2}}}{\sigma_{5{\rm S}_{1/2}}} = 71 \pm 10,
\end{equation}
the error being given only by the uncertainty in $f$.

Also in the ground state 2-photon ionization by the yellow 594 nm
pulsed laser with
photon flux  $F_{\rm ph}\simeq  10^{27}$
cm$^{-2}$s$^{-1}$, we examined the decay of the MOT fluorescence
emission. Using the
method of interleaved cw and synchronously switched runs described
above, we verified that the 2-photon ionization contribution from
the excited 5P$_{3/2}$ state could be neglected in this case. The
additional measured loss rate introduced by the pulsed dye laser
with rate $r=$ 11 Hz, was $\gamma^{\rm eff}_{\rm ph} =1.14 \times
10^{-3}$ s$^{-1}$. Using Eq. (\ref{effrate}) the ionization
probability $P_{\rm ph}$ by a single pulse is $ 1\cdot10^{-4}$,
corresponding to a 2-photon ionization cross section from the
ground state
\begin{equation}
\sigma^{(2)}_{5 {\rm S}_{1/2}} =(11\pm 6)\times 10^{-49} {\rm cm}^4 s.
\end{equation}
This value agrees quite well with the theoretical prediction $5
\cdot 10^{-49}$cm$^4$s, calculated by Bebb\cite{bebb}.

\begin{figure}
\centering \psfig{figure=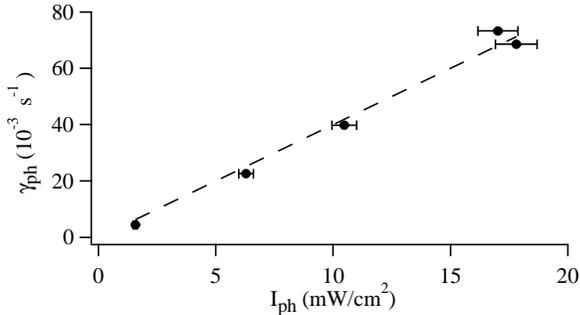, width=8.3cm}
\caption{Additional loss rate $\gamma_{\rm ph}$ introduced in the
MOT evolution by the presence of the 421 nm cw laser, as a
function of the laser intensity $I_{\rm ph}$. From these data the
photoionization cross-section for the excited 5P$_{3/2}$ state was
derived.} \label{fig:2}
\end{figure}
\begin{figure}
\centering \psfig{figure=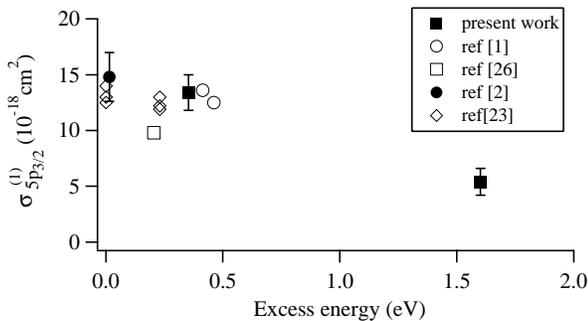, width=8.3cm} \caption{1-photon
ionization cross sections in the Rb MOT from the excited state
5P$_{3/2}$ as a function of the electron excess energy. The filled
squares are our experimental results for $\lambda_{\rm ph} = 296$
nm and $\lambda_{\rm ph} = 421$ nm. Other data points are
extracted from the literature.} \label{fig:3}
\end{figure}
Using the 421 nm cw laser, we studied the decay of the MOT
population due to ionization from the first excited state, whose
cross section was estimated  by Aymar {\it et al} \cite{aymar}.
For the the 2-photon ionization from the ground state the
cross-section $\sigma^{(2)}_{5{\rm S}}= 4\times10^{-49}$ cm$^{4}$s
estimated in\cite{bebb} predicts a negligible ground-state
ionization with the available photon flux.  The measured
additional MOT loss rates $\gamma_{\rm ph}$ are plotted in
Fig.\ref{fig:2} as a function of the photoionizing laser intensity
$I_{\rm ph}$. From these data we obtained the following value for
the photoionization cross-section at $\lambda_{\rm ph}= 421.66$
nm:
\begin{equation}
\sigma^{(1)}_{5{\rm P}_{3/2}}= (1.34 \pm 0.16)\times 10^{-17} {\rm cm}^2
\end{equation}
Figure \ref{fig:3} shows the measured data for the photoionization
cross-section from the rubidium excited state 5P$_{3/2}$ for
various wavelengths of the ionizing laser. Our experimental points
at 296 nm (pulsed) and 421 nm (cw)  are in good agreement with the
values previously measured by Dinneen {\it et al}\cite{dinneen},
Gabbanini {\it et al}
\cite{gabbanini},  and Klyucharev
and Sepman\cite{klyucharev}, the last one for an unresolved fine
structure $5P$ state, and the two theoretical estimates of Aymar
{\it et al}\cite{aymar}.  Within the explored range the
cross-section dependence on $E_{\rm ex}$ is fitted by a
straight line, while at larger excess energy a $E_{\rm ex}^{-9/2}$
dependence
is expected\cite{aymar}.

\begin{figure}
\centering \psfig{figure=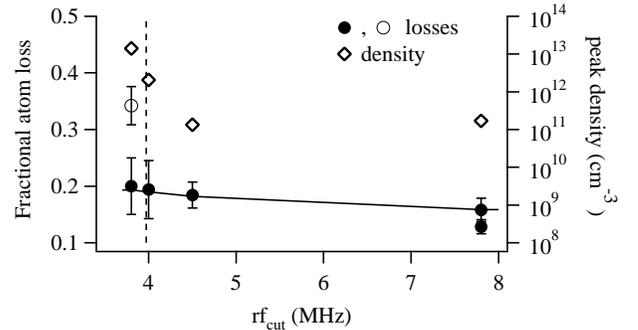,
width=8.3cm}\caption{Fractional loss of atoms after exposure to
120 pulses of the 296 nm laser inside the magnetic trap (filled
circles) as a function of the final frequency for rf evaporation
(for the open circle data point a longer time of flight was used).
The decreasing rf cut produced an increasing atomic peak density
(open diamonds). The dashed line indicates the condensation
threshold position. The continuous line models the effect of the
changing size of the cloud within the spatial distribution of the
uv beam.} \label{fig:7}
\end{figure}

\subsection{Magnetically trapped atoms}
In the ionization experiment performed with uv light at 296 nm a
lower peak intensity (70 $\mu$J) and the large beam size lead to a
single pulse ionization probability of 0.002. Because a single
pulse produced a loss of atoms too small to be directly detected,
we illuminated the magnetically trapped atoms (condensate and
non-condensate) with a sequence of 120 pulses, at a repetition
rate of 11 Hz. We performed the investigation of the ionization
losses starting with an initial cold cloud of $10^6$ atoms and,
varying the final rf frequency cut, we ionized clouds with
decreasing temperature and increasing atom density. Finally, for
radio-frequency cuts below the 3.95 MHz threshold,  we ionized a
pure condensate. Data for the peak density and the fractional atom
loss are reported in Fig.\ref{fig:7} as a function of the final rf
value. The measured total fractional losses between 15 and 20$\%$
were compared to the single pulse loss extracted from the MOT
measurements described by the continuous curve in Fig.\ref{fig:7}
taking into account the changing size of the atomic cloud into the
geometric factor $\beta^{(1)}$ of Eq.(\ref{rate}) and an overall
scale factor of 1.5, due to an estimated misalignement of 100 $\mu
m$ between the atomic cloud and the laser beam. We verified that
the temperatures of the thermal clouds were unaffected by the
laser pulses within the experimental uncertainty of $5\%$.
However, when using a longer time of flight before imaging for a
pure Bose condensed cloud (in order to decrease the optical
density, thus getting more reliable data for the number of atoms),
we measured higher losses (the open circle in Fig. \ref{fig:7}).
This result indicates the presence of very low energy thermal
atoms removed from the condensate but still trapped, albeit with a
spatial density below our detection noise. Thus the 296 nm laser
acting on a condensate produces additional loss of low energy
thermal atoms ejected from the condensate cloud. Since for the
large beam size and the moderate pulse energy we did not expect a
significant contribution of scattering and dipole forces, we have
taken these additional losses as first evidence for the
interaction of ionization products with the condensate background.
\begin{figure}
\centering \psfig{figure=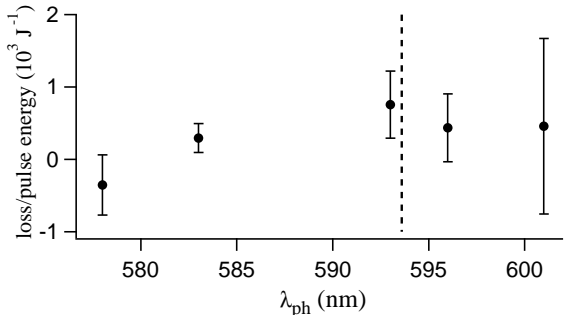, width=8.3cm} \caption{Loss
rate for a single laser pulse, normalized by the pulse energy,  as
a function of the dye-laser wavelength. The vertical line
indicates the threshold wavelength for the 2-photon ionization
from the ground state.} \label{fig:4}
\end{figure}

Using the 594 nm pulsed radiation, we irradiated the magnetically
trapped atoms  either with few pulses or by applying a single
laser pulse. The puzzling result of these investigations was the
measurement of a large loss rate, larger  by a factor ten than the
value expected from the previously measured 2-photon ionization
cross-section. Moreover, the measured trap losses depended
strongly on the alignment of the photoionizing laser with respect
to the magnetic trap. For perfect alignment and small clouds we
expected an ionized fraction of 2$\%$, while losses of 20$\%$ were
observed. An example of the measured fractional loss of atoms from
the magnetic trap, normalized to the laser pulse energy,  is
reported in Fig.\ref{fig:4} as a function of the wavelength of the
ionizing laser. The data of Fig.\ref{fig:4} do not exhibit a clear
evidence of threshold behavior expected for a 2-photon ionization
process. These results suggest the presence of other loss
mechanisms masking the ionizing process. In a magnetic trap, in
fact, every process changing the hyperfine or m-sublevel leads to
atom loss.
\begin{figure}
\centering \psfig{figure=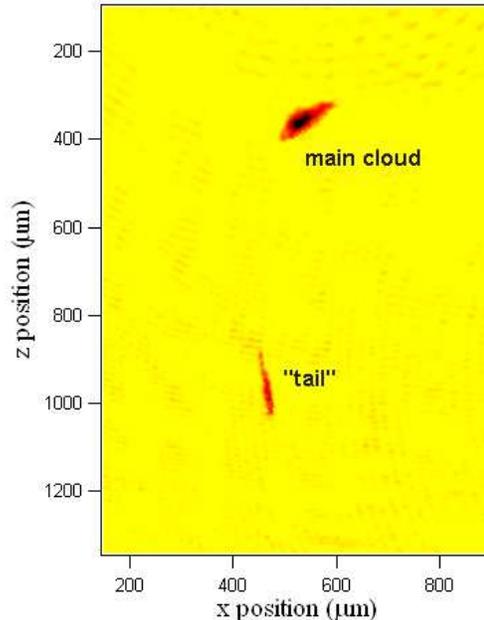, width=7cm} \caption{Shadow
image of a condensate after the application a single dye laser
pulse at 594 nm. The condensate is deformed and a secondary
condensate containing around $30\%$ of the original atoms is
present (the periodic circular fringe patterns are an artifact of
the imaging technique). The size of the main cloud  at the
ionization time was $7 \mu m$, the trapping frequency 23 Hz, the
time of flight 18 ms.} \label{fig:5}
\end{figure}

The presence of more complex trap loss processes is indicated also
by the spatial distribution of the atoms inside the magnetic trap
after the application of a single laser pulse at 594 nm. For both
the condensate and the thermal cloud of magnetically trapped
atoms, we noticed that the laser pulse ejected a secondary cloud,
a "tail", from the original one, as shown in Fig.\ref{fig:5}. The
shadow image of Fig.\ref{fig:5}, obtained after 11 ms of evolution
of the condensates within the magnetic trap, shows two clouds, the
upper cloud at the place of the original one and the ejected
secondary cloud separated spatially, at a lower vertical position.
The secondary cloud of Fig. \ref{fig:5} contained around $30\%$ of
the total number of atoms, but in other conditions the secondary
cloud contained a fraction reaching $50\%$.  Measuring the atomic
density and temperature for the two clouds separately we derived
that the phase space density of the ejected cloud was comparable
to the remaining fraction of the original cloud. Below the
condensation threshold both remaining clouds had a phase space
density far above 1. This suggests that the laser pulse split the
original condensate into two condensed clouds. Both of the
observed clouds were strongly deformed with respect to the
unperturbed condensate showing a close to spherical density
distribution after the same time of flight. The aspect ratio and
orientation of the deformed cloud depended on the evolution time
within the magnetic trap after the laser excitation. Moreover a
detailed analysis, outlined below, showed for the main cloud a
negligible center-of-mass motion, while the secondary cloud
acquired a center-of-mass motion during the interaction with the
laser pulse.

\begin{figure}
\centering \psfig{figure=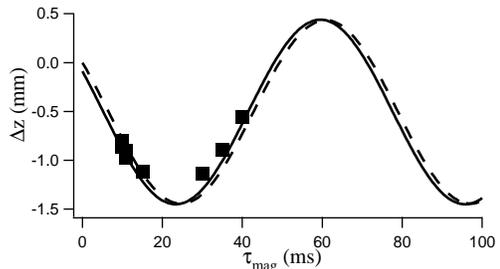, width=7cm} \caption{Data for
the relative displacement $\Delta z$ of the $\mid
F=2,m_F=1\rangle$ secondary condensate, ejected from the $\mid
F=2,m_F=2\rangle$ primary one after a single 594 nm  laser
ionization pulse, versus the evolution time $\tau_{\rm magn}$
within the magnetic trap. The oscillating fit functions correspond
to different initial velocities $v_{\rm z}(0)$ of the secondary
cloud, $v_{\rm z}(0)$=0 for the dashed line and $v_{\rm z}(0)=
-0.5$ cm/s for the continuous line, whose vertical shift for
$\tau_{\rm magn}=0$ is due to free fall evolution of the
condensate during the time-of-flight.} \label{fig:6}
\end{figure}

In order  to understand the nature of the secondary cloud, we used
the magnetic trap as a Stern-Gerlach analyzer.  We measured the
position of the secondary condensate as a function of the
evolution time $\tau_{\rm magn}$ within the magnetic trap after
the application of the laser pulse, with results shown in
Fig.\ref{fig:6}. From the oscillatory behavior of the difference
in position of the center-of-mass for the two clouds we inferred that the secondary
cloud
consisted of rubidium atoms in the $\mid F=2,m_F=1\rangle$ Zeeman
state, excluding the $\mid F=1,m_F=-1\rangle$ state with similar
magnetic moment, because our detection was sensitive only to atoms
in the $F=2$ state\cite{nota}. Owing to the different magnetic
moment of this state, both the frequency of the harmonic sloshing
motion and the equilibrium position differ from  those for atoms
in the $\mid F=2,m_F=2\rangle$ state. A fit to the measured
position data, assuming that the secondary cloud was created at
the instant of the laser pulse from the position of the mother
condensate, revealed that the secondary cloud started its
oscillation with an initial velocity   $v_{\rm
z}(0)$=(-0.5$\pm$0.2) cm/s, i.e., a velocity pointing downwards
along the vertical axis\cite{BTOP}. The laser pulse propagated at
an angle of about $35^{o}$ with respect to the horizontal x-y
place and pointed downwards in the z-y plane, while the laser
light was polarized along the x-direction of Fig.\ref{fig:5}. The
atoms are polarized along the direction of the bias field,
rotating  in the horizontal x-y plane with frequency 10 kHz.
Changing the timing of the laser pulse with respect to the bias
field rotation we found no qualitative difference in the atomic
response for parallel and crossed electric and magnetic fields.

To judge how the direct mechanical effect of the laser pulse could
be responsible of the observed complex behavior of the clouds and
of the anomalously high losses, we estimated the role of the
dipole forces in a single atom picture. Placing a rubidium atom at
the steepest slope of the spatial laser beam profile during  the
pulse, we estimated an atom to acquire a velocity of 0.8 cm/s
under application of the 594 nm pulse. Based on this estimate the
atomic kinetic energy of 40 $\mu$K would be well below the trap
depth, and direct losses due to the dipole force would not be
expected. An additional puzzle is associated with the transfer of
the atoms from the initial $m_{F}=2$ state to the $m_{F}=1$ final
one. Such a transfer could be created by a Raman process
associated to the absorption and emission of the laser photons,
with a probability enhanced by the atomic stimulated emission as
in the atom laser emission. However, the Raman transfer between the
Zeeman sublevels is forbidden by the $\Delta m_F=0$ selection rule
for the interaction with a linearly polarized laser field. A
violation of the selection rule may arise because the trap
magnetic field is not constant during the laser pulse or by
fluctuations of the laser polarization on the time scale of a
single pulse. Notice that Raman scattering processes do not
produce losses in a MOT, where every Zeeman sublevel of the ground
state is trapped and  the kinetic energy acquired is smaller than
the trap depth. As mentioned above, the shape of the detected
clouds was grossly different from the shape of unperturbed
condensates. Such a shape deformation could be associated with the
asymmetry in the electron emission produced by the photoionization
process, exciting complicated oscillations in the original and
secondary condensate clouds.

Finally we excited thermal clouds and condensates with the 421 nm
radiation detuned by 12 GHz above the resonance and focused to
$30\mu m \times 3.3 \mu m$. At such a detuning the scattering
force still dominates the atomic losses from the magnetic trap,
and the trap losses due to the 2-photon ionization from the ground
state could not be directly measured.  We observed that the
condensate cloud falling under gravity through the 421 nm laser
beam was split into two condensate clouds, the reflected and
transmitted respectively, again without measurable additional
atomic losses. Note that in this case the interaction time with
the laser radiation is orders of magnitude longer than for
pulsed lasers, so that a complicated dynamics may be produced.

\section{Conclusions}
We applied trap loss spectroscopy to measure the photoionization
cross-sections of rubidium atoms confined in a MOT making use of
cw and pulsed laser. We verified that the use of pulsed lasers did
not limit the accuracy reached in the cross-section
determination. We demonstrated that the photoionization of
magnetically trapped thermal and condensed atoms instead does not
represent a precise tool for cross-section determination. In
effect the magnetic trap depth is smaller than the MOT depth, and
the laser action on the cold atomic cloud may produce additional
processes leading to atomic losses from the trap, masking the
ionization losses. We have observed that both scattering and
dipole forces associated with strong pulsed lasers may impart a
kinetic energy large enough to overcome the trap depth. Moreover,
dealing with a  condensate cloud, additional losses appear that
are probably due to the creation of a dilute thermal cloud.

We have discovered that the application of intense pulsed
radiation leads to huge excitation and macroscopic splitting of a
condensate without compromising its phase space density. We
identified the transfer of atoms to the $\mid F=2, m_F=1 \rangle$
level as an important process, but the detailed microscopic
mechanism remains unclear and will be subject of further studies.

On the basis of the photoionization cross-sections measured within
a MOT we estimated to have created charged clouds and condensates
in magnetic traps containing a fraction of 0.2$\%$ and 2$\%$ of
ions at a time for the 1-photon and 2-photon experiments
respectively. We have not observed any dramatic instability which
could be triggered by the strong polarization of those charged
condensates. However, we take the higher losses measured following
the 1-photon ionization of a condensate as a first indication of
the interaction between the ionization products and the condensate
background. Additional processes such as collisions with cold
electrons and ultra-cold neutrals, recombination and collective
processes in the weak and cold plasma could modify the atomic
cloud evolution and in part be responsible for the more
complicated dynamics observed after the 2-photon excitation in the
condensate. In future experiments we plan to use cw sources to
ionize condensed clouds to disentangle scattering and dipole force
effects from the rich and interesting physics of cold charged
particles in quantum degenerate clouds.

\section{Acknowledgments}
The authors wish to thank G. Alber, W.M. Fairbank Jr., C. Fort, P.
Gould, M. Inguscio, and P.
Zoller for illuminating discussion. This research was supported by the
Sezione A of INFM-Italy through a PAIS Project,  by the
MIUR-Italy through a PRIN Project, and by the EU
through the contract HPRN-CT-2000-00125.


\begin{references}
\bibitem{helium} In the $He^*$ condensates of \cite{helium1}
ions were present due to Penning collisions, but
the interaction time between ions and  neutrals was extremely short
due to the high realesed energy.
\bibitem{nota}We confirmed the absence of atoms in the $|F=1 \rangle$
hyperfine states by applying a repumper laser before imaging the
primary and secondary clouds and verifying that the atom numbers
in the clouds were not modified.
\bibitem{BTOP}We allowed for a 2$\%$ accuracy in the magnetic
field calibration.
\end{references}
\end{document}